\documentclass[sigconf,nonacm]{acmart}

\usepackage{booktabs}
\usepackage{graphicx}
\usepackage{amsmath}
\usepackage{tikz}
\usetikzlibrary{arrows.meta,positioning,fit,backgrounds}
\usepackage[capitalise]{cleveref}

\setcopyright{none}
\renewcommand\footnotetextcopyrightpermission[1]{}
\begin{document}

\title{Stateless Network-Aware Adaptive Bitrate Streaming over IPFS}

\author{Iliya Mirzaei}
\authornote{These authors contributed equally to this work.}
\affiliation{%
  \city{Stony Brook}\state{NY}\country{USA}}
\email{imirzaei@cs.stonybrook.edu}
\author{Shabnam Jafarzade Mojaveri}
\authornotemark[1]
\affiliation{%
  \city{Stony Brook}\state{NY}\country{USA}}
\email{sjafarzademo@cs.stonybrook.edu}
\author{Amirhossein Najafizadeh}
\authornotemark[1]
\affiliation{%
  \city{Stony Brook}\state{NY}\country{USA}}
\email{anajafizadeh@cs.stonybrook.edu}

\begin{abstract}

Modern content delivery systems are increasingly adopting
decentralized architectures to improve availability,
reduce operational costs, and better serve geographically
distributed users.
The InterPlanetary File System (IPFS) represents a
promising approach by using content-based identifiers
distributed across a global peer-to-peer network.
Although IPFS improves performance, fault tolerance,
resilience, and censorship resistance, its inherently
unpredictable environment introduces significant performance
variability that limits the effectiveness of conventional
Adaptive Bitrate (ABR) streaming mechanisms and degrades
Quality of Experience (QoE).
Recent network-aware ABR solutions have emerged to address
this challenge by incorporating IPFS-specific information
into bitrate selection decisions.
However, these approaches rely on maintaining continuously
synchronized state across consumers and providers during
content delivery, which can quickly become stale in the
presence of peer churn, provider migrations, network
partitions, and changing content distributions.
As a result, existing network-aware ABR policies become
less effective.
In this paper, we investigate whether network-aware ABR
approaches can remain effective without maintaining synchronized
adaptation state. To this end, we present a
\textit{stateless network-aware ABR} policy for
IPFS-based video streaming.
Our approach replaces provider-stateful adaptation with an
observation-driven policy that recomputes bitrate decisions
for each segment using only locally observable request-time signals.
To preserve end-to-end adaptation context without maintaining
synchronized state at providers, the client embeds its adaptation
state in HTTP headers, ensuring that state remains under
client control and is carried transparently across requests.
By eliminating cross-provider state synchronization,
the framework improves robustness to system
failures and network reconfigurations while simplifying
deployment at scale.
Early results show the approach achieves up to roughly $6\times$
higher QoE than a non-adaptive, uncached configuration in our testbed.
These findings demonstrate that stateless network-aware adaptation
provides a practical and scalable foundation for decentralized video delivery.

\end{abstract}


\maketitle

\section{Introduction}

Content delivery is shifting from centralized origins toward decentralized
architectures that improve availability, lower operational cost,
and place content closer to geographically distributed users
~\cite{vu2019blockchain, kim2018ddos, goyal2019secure}.
Recent years have seen growing interest in decentralized content
delivery platforms such as the InterPlanetary File System
(IPFS)~\cite{ipfs, benet2014ipfs}.
Such systems are particularly attractive in environments where
centralized platforms are unavailable, restricted, or difficult to
trust. Studies of alternative software distribution ecosystems show
that users frequently rely on decentralized or third-party content
sources when access to official distribution channels is limited
~\cite{khanlari2026forbidden}.

%
%

IPFS replaces traditional location-based addressing with cryptographic content
identifiers (CIDs) and distributes content across a global peer-to-peer network.
Content can be replicated and served by multiple independent providers.
Therefore, users can retrieve data from any reachable replica rather than
a single origin server~\cite{ipfs}.
This architecture improves availability, fault tolerance,
and geographic reach while reducing dependence on centralized
infrastructure~\cite{ipfs}.

%
%

Video streaming is among the most demanding content delivery workloads,
driving the need for highly scalable and geographically
distributed content delivery infrastructures~\cite{qiao2025pira}.
The decentralized design of IPFS offers an appealing alternative
to traditional CDN systems by enabling content replication across a
global peer-to-peer network without relying on
centralized servers~\cite{minegishi2025video}.
In principle, these properties can improve availability,
reduce infrastructure costs, and increase resilience
to failures~\cite{qiao2025pira}.
In practice, however, they introduce substantial variability into
the content delivery process~\cite{telescope2023}.
Since content may be served by different providers and traverse
different network paths from one request to the next,
the performance seen by a streaming client becomes significantly
less predictable than in conventional CDN systems.

State-of-the-art content video streaming systems use HTTP-adaptive bitrate (ABR)
streaming which encodes content at several bitrates and the client continually selects
the highest bitrate it believes the network can sustain~\cite{dashif}.
ABR algorithms highly depend on a stable and predictable estimate of available 
throughput~\cite{bola2016, mpc2015, bba2014, pensieve2017}.
With traditional CDN, consecutive video segments are often served
from the same cache or a small set of nearby servers, making past
throughput measurements useful predictors of future performance~\cite{tyagi2025optimizing}.
In IPFS, however, each segment request may be satisfied by a different provider,
traverse a different network path, or experience a different lookup delay.
Many factors such as host availability, content placement,
and content retrieval paths can change rapidly~\cite{trautwein2022},
causing large fluctuations in latency and throughput.
As a result, a throughput sample observed for one segment may be a poor
predictor of the next, leading conventional ABR algorithms to oscillate
between bitrates or incur playback stalls.

Network-aware ABR systems have emerged to address this
challenge~\cite{telescope2023, bhat2018sabr, palit2023improving}.
These systems improve Quality of Experience (QoE) by incorporating
specific network information into bitrate selection.
However, they rely on maintaining state, such as throughput histories
and provider performance statistics, across clients,
providers, and intermediary proxies.
In decentralized environments, maintaining such state is
inherently difficult~\cite{he2026post}.
Peer churn, provider migration, network partitions, and shifting content placement
can quickly invalidate previously collected information, causing adaptation decisions
to be based on stale observations. Moreover, synchronizing state across distributed
components introduces additional complexity and creates scalability bottlenecks
as the number of clients, providers, and proxy instances grows.
This raises a fundamental question:
\emph{Can network-aware ABR remain effective in IPFS without relying on maintained state?}

To answer this question, we present a soft-stateless network-aware ABR policy
for IPFS video streaming. Rather than maintaining long-lived synchronized state,
our approach makes bitrate decisions independently for each segment using only
information available at request time, including recent segment throughput,
cache hit or miss status, gateway and node response latency, and client buffer
occupancy. Any state retained is strictly local and short-lived, existing only
long enough to inform the current adaptation decision.
Early results demonstrate that the proposed approach achieves up to
roughly a 6$\times$ higher QoE than a non-adaptive, uncached configuration
in our testbed.





\section{Background}
\label{sec:bg}

\paragraph{Decentralized Content Delivery}
IPFS stores data as a Merkle DAG of content-addressed blocks~\cite{sanjuan2020merkle}.
Each block is identified by a Cryptographic Content Identifier (CCID) derived from its hash.
Peers locate content providers through a Kademlia-style Distributed Hash Table (DHT)~\cite{kademlia2002}.
Kubo~\cite{kubo}, the reference implementation of IPFS, exposes an HTTP gateway that allows
conventional web clients to retrieve content without native IPFS support.
A gateway request may be satisfied from a local block store,
retrieved from a directly connected peer through Bitswap, or resolved
through DHT-based provider discovery
~\citep{balduf2022monitoring, delarocha2021bitswap}.
Unlike traditional content delivery infrastructures, IPFS operates in
an environment where content availability and retrieval paths can vary
significantly over time. The same object may be replicated across
different providers, served through different network paths, or become
temporarily unavailable as peers join and leave the network.
Measurements have shown that retrieval latency and throughput vary
widely depending on content popularity, replication levels, provider availability,
and peer churn ~\cite{trautwein2022,daniel2022ipfs,daniel2022passively,shi2024closer}.

\paragraph{Adaptive Bitrate Streaming}
Protocols, such as DASH~\cite{dashif}, segments video content and encodes each segment
at multiple bitrate representations. A Media Presentation Description
(MPD) specifies the available representations, while clients
dynamically select a bitrate for each segment during playback~\cite{dashjs}.
ABR algorithms differ primarily in the signals they use to estimate
future network conditions. Rate-based approaches track recent
throughput measurements, buffer-based approaches such as
BOLA~\cite{bola2016} and BBA~\cite{bba2014} rely on playback buffer
occupancy, control-theoretic approaches such as MPC~\cite{mpc2015}
optimize decisions over a short prediction horizon, and learning-based
approaches such as Pensieve~\cite{pensieve2017} learn adaptation
policies directly from observed network behavior.
Early studies demonstrated that ABR clients can react slowly to
bandwidth changes, oscillate between bitrate levels, and behave
unfairly when competing for shared network resources
~\citep{akhshabi2011experimental,jiang2012festive}.
Subsequent surveys and modern ABR systems
~\citep{bentaleb2019survey,chen2024soda}
continue to balance three competing objectives: maximizing video
quality, minimizing re-buffering events, and reducing quality
oscillations. Achieving this balance relies on the assumption that
recent network observations remain predictive of near-future delivery
conditions.

\paragraph{Network-Aware ABR}
Several systems have proposed incorporating IPFS-specific network
information into bitrate adaptation decisions. Telescope
~\cite{telescope2023}, the closest prior work, places a proxy between
the IPFS gateway and the client, maintains throughput histories for
both gateway and network retrieval paths, and rewrites the MPD so that
the client can adapt to observed IPFS retrieval conditions.
Experimental results demonstrate that exposing network-level retrieval
information improves Quality of Experience (QoE), measured through a
combination of average video quality, re-buffering events, and quality
stability~\cite{mok2011,dobrian2011}.
More broadly, data-driven ABR systems such as
CS2P~\cite{sun2016cs2p} improve adaptation decisions through
historical throughput observations and prediction models.
Network-assisted adaptation has also been explored in other dynamic
environments. For example, SARA~\cite{fang2024robust} incorporates
predictions of connectivity disruptions in LEO satellite networks into
bitrate adaptation decisions to reduce re-buffering and quality
degradation.
These systems demonstrate that exposing additional network information
can improve bitrate selection under challenging network conditions.
Existing IPFS-aware ABR systems improve adaptation by maintaining
historical information about network and provider behavior.

\section{Stateless Network-Aware ABR Design}
\label{sec:design}

\begin{figure}[t]
\centering
\resizebox{\columnwidth}{!}{%
\begin{tikzpicture}[
  font=\small,
  node distance=8mm and 10mm,
  >={Stealth[length=2mm]},
  ext/.style={draw, rounded corners, minimum height=9mm, minimum width=16mm, align=center, fill=blue!6},
  mod/.style={draw, rounded corners, minimum height=8mm, minimum width=22mm, align=center, fill=orange!12},
  obs/.style={draw, rounded corners, minimum height=9mm, minimum width=44mm, align=center, fill=green!10}
]
  \node[ext] (client) {DASH.js\\client};

  \node[mod, right=28mm of client] (router) {Request router\\(Fiber)};
  \node[mod, above=6mm of router] (estim) {Bandwidth estimator\\$T_c,\,T_g,\,T_n$};
  \node[mod, right=12mm of estim] (rewrite) {MPD rewriter};
  \node[mod, below=6mm of router] (cache) {Two-layer cache\\(memory $\to$ file)};
  \node[mod, right=12mm of router] (abr) {ABR selector\\(throughput /\\statistics)};

  \node[ext, right=28mm of cache] (ipfs) {Kubo IPFS\\cluster\\(3 nodes)};

  \node[obs, below=16mm of cache] (obsv) {OpenTelemetry $\to$ Prometheus + Jaeger};

  \draw[<->] (client) -- node[above,font=\scriptsize]{HTTP} (router);
  \draw[->] (router) -- (cache);
  \draw[->] (cache) -- node[above,font=\scriptsize]{CID GET} (ipfs);
  \draw[->] (router) -- (estim);
  \draw[->] (estim) -- (rewrite);
  \draw[->] (router) -- (abr);
  \draw[->] (abr.north) -- (rewrite.south);

  \draw[->,dashed] (cache) -- (obsv);
  \draw[->,dashed] (router.south) to[bend right=15] (obsv.north);

  \begin{scope}[on background layer]
    \node[draw, dashed, rounded corners, fill=gray!4,
          fit=(router)(cache)(estim)(rewrite)(abr),
          inner sep=5mm, label=above:{\textbf{Telescope proxy (stateless)}}] (proxybox) {};
  \end{scope}
\end{tikzpicture}%
}
\caption{Architecture of the re-engineered Telescope proxy. The stateless proxy mediates between the DASH.js client and a Kubo IPFS cluster, serving segments from a two-layer cache.}
\label{fig:arch}
\end{figure}
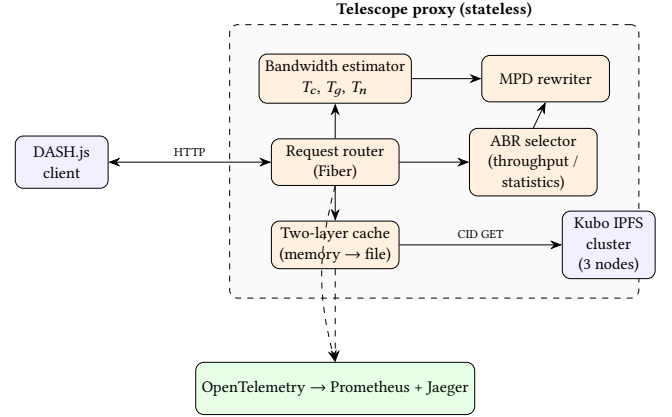

Our design pursues a different approach. Each bitrate decision is
derived from observations available at request time rather than from
long-lived adaptation state.
The resulting design preserves network-awareness while avoiding
synchronized throughput histories, provider-performance tables, or
other shared adaptation metadata.
Figure~\ref{fig:arch} illustrates the proposed architecture.
As shown in the Figure~\ref{fig:arch},
the system is deployed as a transparent proxy between a
streaming client and a IPFS gateway.
For each video request, the proxy observes current retrieval
conditions, estimates the bandwidth available to the client, and
rewrites the MPD to expose only representations that can be sustained
under those conditions.
The client then continues to use its native ABR algorithm on the
modified representation ladder.
Because each decision is made independently from current observations,
any proxy replica can serve any request without coordinating adaptation
state with other replicas.

The proxy estimates available bandwidth using three runtime signals:
(i) client throughput, $T_c$, measured as the recent segment delivery
rate observed by the player,
(ii) gateway fetch latency, $T_g$, representing the time required to
retrieve content from the IPFS gateway, and
(iii) IPFS node throughput, $T_n$, capturing the bandwidth of recent
network retrievals from the IPFS network.
The proxy additionally observes whether the requested segment is
available in the local cache.
Using these signals, the proxy derives an estimate of the bandwidth
available for video delivery and removes bitrate representations that
cannot be sustained under current conditions.
This allows the client ABR algorithm to begin adaptation from a
realistic representation set rather than discovering bandwidth
limitations through playback stalls.

We implement two alternative adaptation strategies.
The \emph{throughput-based} strategy estimates the usable video
bandwidth by discounting the IPFS-related retrieval overhead from the
observed client throughput:
\begin{align}
\text{Bandwidth} &= T_c - T_g & &\text{(cache hit)} \label{eq:tp-cached}\\
\text{Bandwidth} &= T_c - T_n & &\text{(cache miss)} \label{eq:tp-uncached}
\end{align}
The \emph{statistics-based} strategy combines cached and uncached
bandwidth estimates to account for the possibility that future segment
requests may encounter different cache states:
\begin{align}
\text{Bandwidth} &= \tfrac{1}{3}\,\text{cached} + \tfrac{2}{3}\,\text{uncached} & &\text{(cache hit)} \label{eq:stat-cached}\\
\text{Bandwidth} &= \tfrac{1}{3}\,\text{uncached} + \tfrac{2}{3}\,\text{cached} & &\text{(cache miss)} \label{eq:stat-uncached}
\end{align}

The weighting factors are inherited from the reference
implementation and are not tuned in this work.

Since IPFS retrieval latency can significantly affect playback
performance, each proxy maintains a two-tier cache consisting of an
in-memory cache and a file-backed cache.
Requests are resolved from memory first, then disk, and finally from
the IPFS network.
The cache serves purely as a retrieval optimization and is independent
of the adaptation logic.
Consequently, cache contents may differ across replicas without
affecting the correctness of adaptation decisions.

The prototype is implemented as a lightweight Go service positioned
between the DASH client and the Kubo gateway\footnote{\url{https://github.com/iliyami/telescope}}.
The proxy processes client requests, gateway interactions, cache
lookups, and telemetry collection through independent concurrent
execution pipelines implemented using Go's native concurrency
primitives. Efficient communication between these components is
performed through Go channels, which provide lightweight
synchronization and message passing mechanisms~\cite{amir2023}.
To support evaluation and debugging, the system exports telemetry
through OpenTelemetry, Prometheus, and Jaeger
~\cite{opentelemetry,prometheus,jaeger}.

\section{Experimental Methodology}

\paragraph{Testbed.}
We deploy a three-node Kubo~\cite{kubo} IPFS cluster with Docker Compose; all
services run as containers with no network emulation. The content is the Big
Buck Bunny reference video at four content sizes (5, 10, 20, and 30\,MB), each
encoded with \texttt{ffmpeg}~\cite{ffmpeg} into a five-rung bitrate ladder
(426p--4K at 1, 3, 6, 9, and 12\,Mbps) and segmented into more than 200 chunks;
a bootstrap service uploads the segments to IPFS and records their CIDs, and a
DASH.js client streams through the proxy.

\paragraph{QoE metric.}
Following the original formulation~\cite{telescope2023}, we score QoE per
playback as
\begin{equation}\label{eq:qoe}
\mathrm{QoE} = \frac{1}{N}\sum_{n=1}^{N} q_n \;-\; \mu\,\mathrm{StallRate}
\;-\; \mathrm{QualityVariation},
\end{equation}
where $q_n$ is the bitrate (Mbps) of segment $n$ over the $N$ segments of a
playback, the stall rate is (total playback time $-$ video length)$/$video
length, the quality variation is the mean absolute bitrate change between
consecutive segments, and $\mu = 6$. QoE is thus in Mbps-equivalent units
bounded above by the top ladder rate (12\,Mbps), and we use it to compare
configurations within the study.

We sweep a full factorial of two adaptation strategies (throughput- and
statistics-based), three caching policies (none, in-memory, and file-based), and
three proxy replication levels (one to three replicas), giving 18 configurations;
each is run 5--10 times and the reported values are averages. Network-level
quantities (IPFS bandwidth, round-trip time, and hop count) are observed per run
rather than controlled, so any relationship we report between them and QoE is
correlational. We organize the evaluation around four questions: (RQ1) how much,
and through which tier, does caching affect QoE and stalls; (RQ2) does
throughput- or statistics-based adaptation perform better; (RQ3) does proxy
replication improve QoE, and finally (RQ4) using the existing metrics,
can we predict the QoE for decisions?

\section{Results}
\label{sec:results}

We report results per research question. All values are read from the measured
figures, and we compare configurations within the study rather than against an
external baseline.

Across all experiments, a consistent pattern emerges: client-perceived quality
is governed primarily by content availability rather than by network topology or
proxy deployment structure. Configurations that increase cache effectiveness
consistently reduce stalls and improve QoE, while changes to adaptation
strategy and replication level have comparatively smaller effects. The results
therefore suggest a hierarchy of influence on streaming performance: caching is
the dominant factor, adaptation strategy provides secondary gains when cache
hits are unavailable, and replication mainly improves resilience and fetch-path
diversity rather than user-visible quality. This observation is significant for
stateless adaptation because the strongest predictors of QoE—recent throughput
and cache availability—are both observable at request time and require no
synchronized adaptation state across clients, providers, or proxy replicas.

\subsection{RQ1: Caching}
\begin{figure}[t]
  \centering
  \includegraphics[width=\columnwidth]{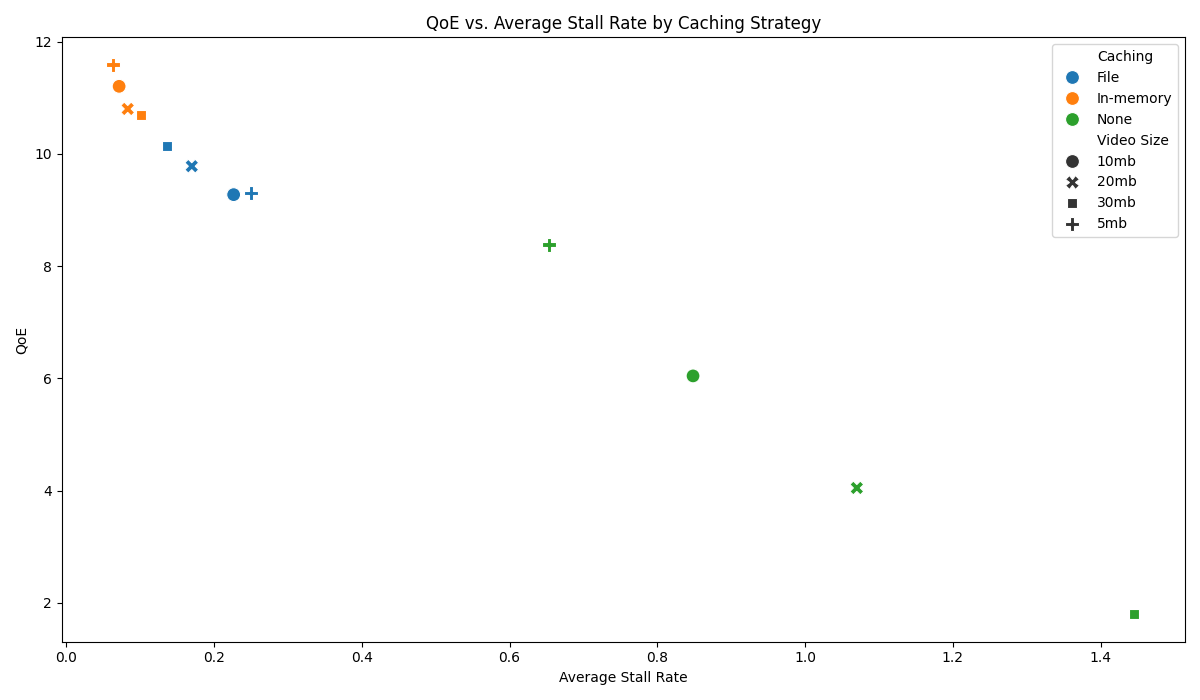}
  \caption{QoE versus average stall rate by caching strategy.}
  \label{fig:qoe-stall}
\end{figure}

Caching is the dominant QoE lever. \Cref{fig:qoe-stall} shows QoE rising sharply
as the cache absorbs retrieval latency: in-memory caching clusters at low stall
rate ($\approx 0.05$--$0.10$) and high QoE ($\approx 10.7$--$11.6$), file-based
caching sits close behind ($\approx 9.2$--$10.2$), and no-caching spreads to high
stall rates ($\approx 0.65$--$1.45$) and low QoE ($\approx 1.8$--$8.4$). The gap
widens with content size: for the largest clip, QoE falls from $\approx 11.6$
with in-memory caching to $1.80$ with none, a roughly $6\times$ difference, and
telemetry recorded cache hit rates rising by up to $\approx 60\%$ under layered
caching. In-memory yields the best QoE but is bounded by RAM, so file-based
caching is the practical default, trading a small QoE reduction for durability
and capacity.

\subsection{RQ2: Adaptation strategy}
\begin{figure}[t]
  \centering
  \includegraphics[width=\columnwidth]{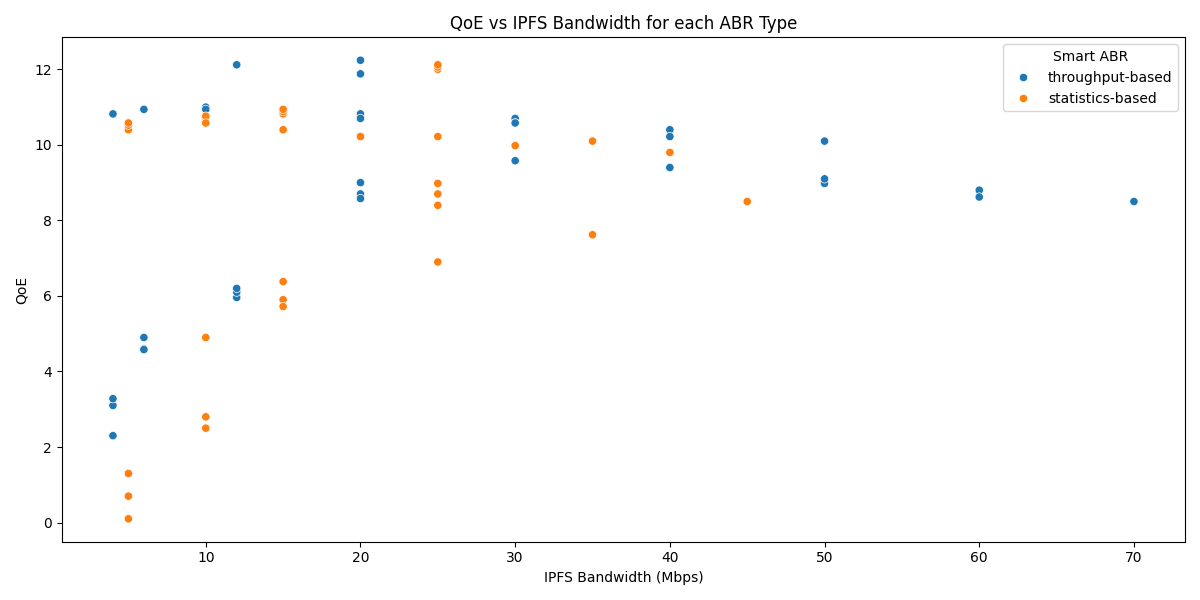}
  \caption{QoE versus IPFS bandwidth across adaptation strategies.}
  \label{fig:qoe-ipfsbw}
\end{figure}

\begin{figure}[t]
  \centering
  \includegraphics[width=\columnwidth]{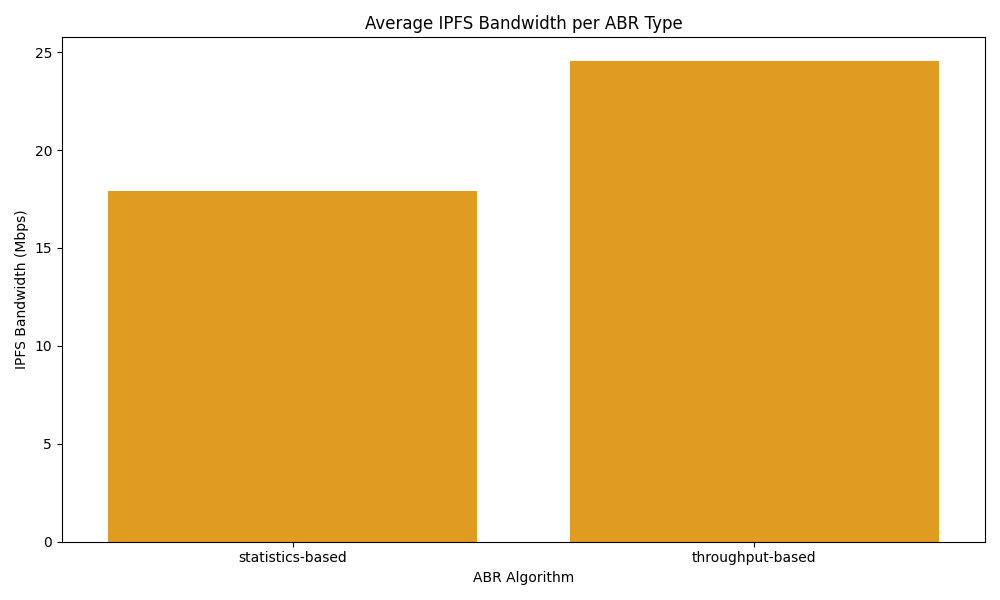}
  \caption{Average IPFS bandwidth per adaptation strategy.}
  \label{fig:ipfs-bw-abr}
\end{figure}

Throughput-based adaptation is the more robust default. \Cref{fig:qoe-ipfsbw}
shows QoE rising with available IPFS bandwidth and saturating around $10$--$12$
for both strategies, but the throughput-based strategy makes fuller use of
capacity: it attains higher average IPFS bandwidth ($\approx 24.5$ versus
$\approx 18$\,Mbps, \cref{fig:ipfs-bw-abr}) and higher gateway bandwidth
($\approx 540$ versus $\approx 447$\,Mbps). Its advantage is largest without
caching, where raw retrieval latency dominates; once a cache absorbs that
latency the two strategies converge (in-memory QoE $\approx 11$, file-based
$\approx 9$--$10$ for both).

\subsection{RQ3: Replication}
\begin{figure}[t]
  \centering
  \includegraphics[width=\columnwidth]{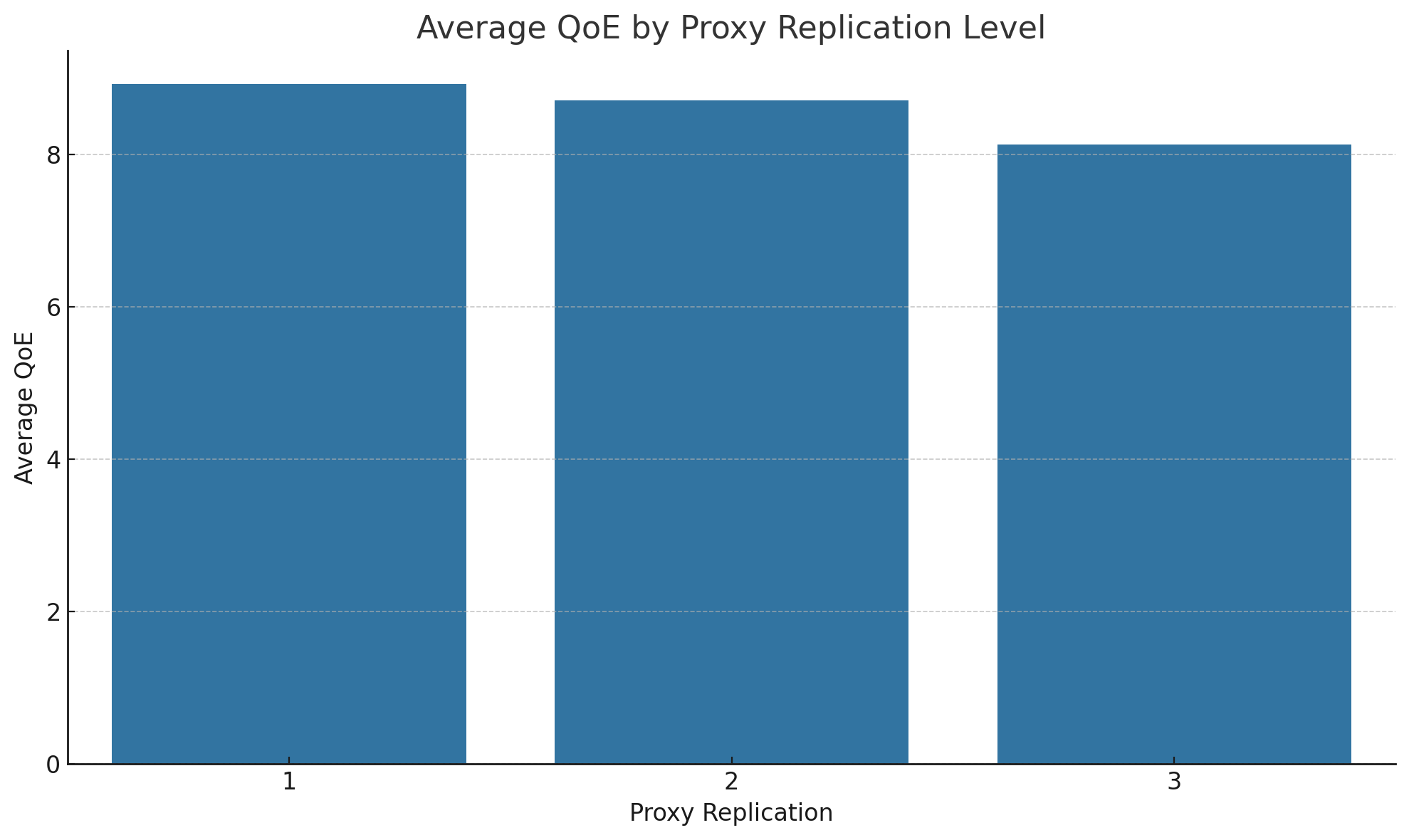}
  \caption{Average QoE by proxy replication level.}
  \label{fig:qoe-replication}
\end{figure}

Replication does not raise QoE. \Cref{fig:qoe-replication} shows average QoE
decreasing slightly from $\approx 8.9$ at one replica to $\approx 8.1$ at three,
a change on the order of the run-to-run variability. Because each stateless
replica holds an independent cache, spreading requests across replicas lowers
per-replica hit rates; we did not measure per-replica hit rates directly, so we
offer cache dilution as a conjecture rather than a confirmed mechanism.
Replication still reduced average fetch latency in congested, high-hop runs (by
up to $\approx 45\%$ in telemetry), so it aids resilience even when QoE gains are
marginal. For QoE-oriented scale-out a shared caching tier is therefore
preferable to more independent replicas.

\subsection{RQ4: What predicts QoE}
\begin{figure*}[t]
  \centering
  \includegraphics[width=0.71\textwidth]{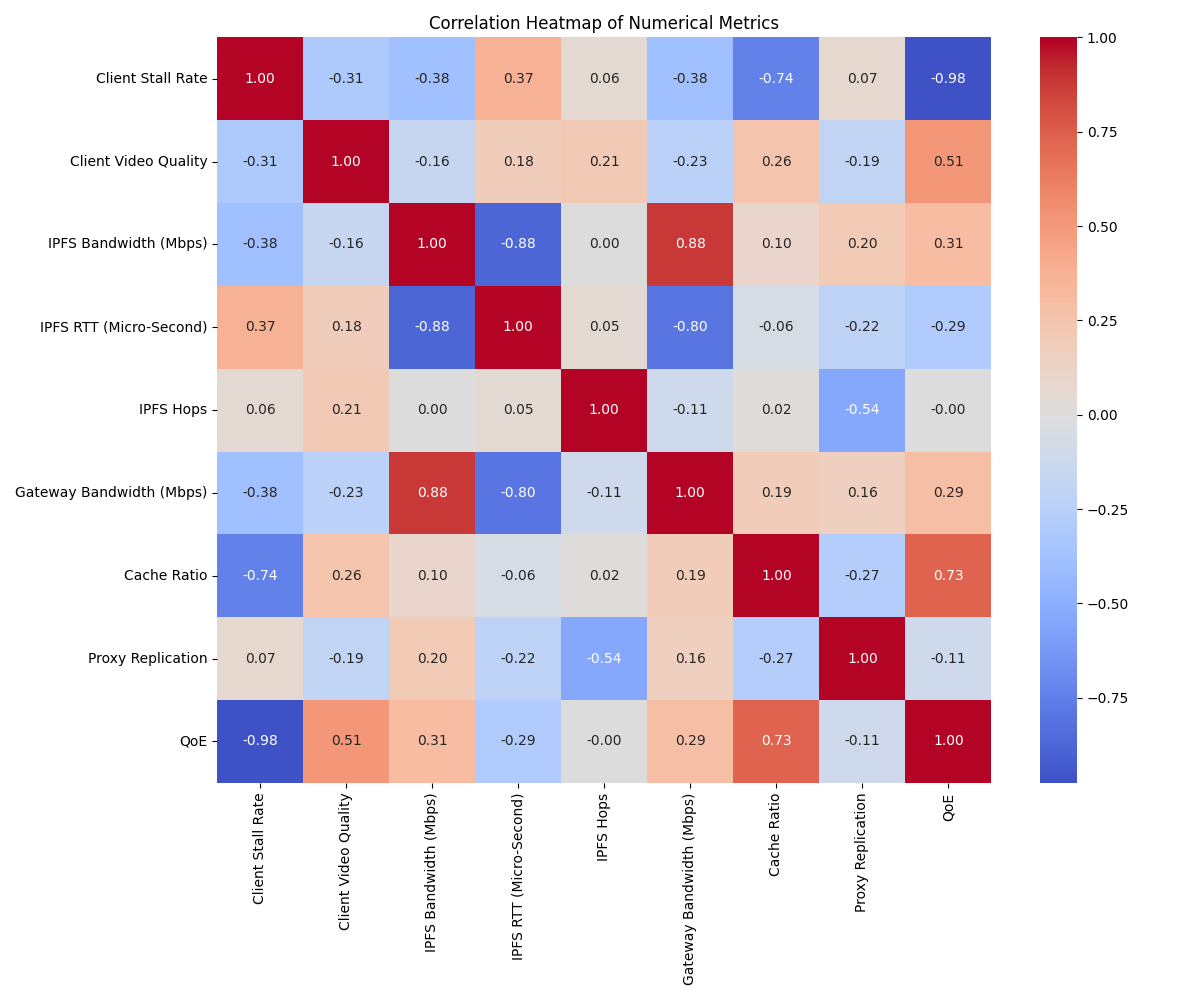}
  \caption{Correlation heatmap of the measured numerical metrics.}
  \label{fig:corr}
\end{figure*}

\Cref{fig:corr} summarises pairwise correlations among the measured metrics.
Stall rate is almost perfectly inversely correlated with QoE ($r = -0.98$) and
bitrate positively so ($r = +0.51$); both are partly definitional, as they are
terms of \cref{eq:qoe}. The informative result is that cache ratio, which is
\emph{not} a term of the metric, is strongly correlated with QoE ($r = +0.73$),
whereas proxy replication ($r = -0.11$) and IPFS hop count ($r \approx 0$) are
essentially uncorrelated. Caching, not topology, governs client-perceived
quality, reinforcing RQ1 and RQ3.

\section{Discussion}

\paragraph{Conclusion}
This paper argues that maintaining synchronized adaptation state is a poor fit
for network-aware ABR over IPFS, where provider churn, content redistribution,
and changing retrieval paths can quickly invalidate stored information.
We presented a soft-stateless design that recomputes bitrate decisions from
request-time observations alone, requiring no shared adaptation state across
clients, providers, or proxy replicas. Experimental results on a real Kubo
testbed show that the design sustains high QoE, achieving up to roughly
$6\times$ the QoE of a non-adaptive, uncached configuration. Across all
evaluated configurations, caching proved to be the dominant factor affecting
QoE, throughput-based adaptation emerged as the most robust default strategy,
and proxy replication improved resilience but did not improve QoE. These
findings suggest that stateless adaptation is a practical foundation for
decentralized video delivery systems.

In this paper we characterize configurations of our system
relative to one another and includes \emph{no external baseline}, neither the
original Telescope nor a non-IPFS-aware client, so we make no claim of
improvement over prior systems. More fundamentally, the robustness
and scalability benefits we attribute to statelessness are architectural
arguments, not measured outcomes.
Taken together, the results indicate that QoE in IPFS-based streaming is driven
more by retrieval efficiency than by network structure. Caching consistently
improves content availability and reduces stalls, throughput-based adaptation
extracts additional performance when retrieval latency is exposed, and
replication primarily improves resilience rather than viewing quality. The
strong correlation between cache ratio and QoE, combined with the weak
correlation of replication and hop count, suggests that maintaining complex
distributed adaptation state provides limited practical benefit. Instead,
high-quality streaming can be achieved largely through local observations and
efficient content reuse, supporting the feasibility of stateless network-aware
ABR in decentralized environments.

\paragraph{Future Work}
Several promising directions remain for future work.
A direct comparison with stateful network-aware ABR systems,
such as Telescope, would help quantify the resilience and scalability
advantages of stateless adaptation under provider churn,
failures, network partitions, and large-scale deployments.
Extending the evaluation to larger and more geographically
distributed IPFS environments would further validate the generality
of the approach. Future research can also explore stateless adaptation
policies specifically designed for request-time observations,
including lightweight prediction and learning-based techniques.
Finally, our results identify cache dilution as the primary limitation
of independent replica scaling, motivating shared-content caching
architectures that preserve stateless operation while improving
cache efficiency and overall system performance.
Collectively, these directions offer a path toward scalable,
fault-tolerant, and fully decentralized video streaming systems built on IPFS.

\bibliographystyle{ACM-Reference-Format}
\bibliography{references}

\end{document}